\documentclass[preprint, notitlepage, longbibliography]{revtex4-2}

\usepackage{amsmath}
\usepackage{amssymb}
\usepackage{bm}
\usepackage{epsfig}
\usepackage{graphicx}
\usepackage{color}
\usepackage[colorlinks]{hyperref}
\usepackage{dcolumn}
\usepackage{mathtools}
\usepackage{gensymb}
\usepackage[displaymath, mathlines]{lineno}

\def\vec#1{{\bf#1}}

\begin{document}

\title{Atomic displacements drive flat band formation and lateral electron and hole separation in near-60$^\circ$ twisted MoSe\texorpdfstring{$_2$}{2}/WSe\texorpdfstring{$_2$}{2} bilayers}

\author{Madeleine Phillips and C. Stephen Hellberg}
\affiliation{Naval Research Laboratory, Center for Computational Material Science, Washington DC 20375, USA }
\affiliation{*Corresponding author email: madeleine.phillips2.civ@us.navy.mil}

\begin{abstract}
	Transition metal dichalcogenide (TMD) bilayers with an interlayer twist exhibit a moir\'e super-period, whose effects can manifest in both structural and electronic properties. Atomic displacements can lead to reconstruction into domains of aligned stacking, and flat bands can form that may host correlated electron states.  In heterobilayers angular mismatch is nearly unavoidable, so understanding the consequences of an interlayer twist is essential.  Using \emph{ab initio} density functional theory, we find that in near-60$^\circ$ twisted MoSe$_2$/WSe$_2$ bilayers valence and conduction band flat bands emerge at $\sim3^\circ$ twist.  Despite relatively limited reconstruction at these angles, atomic displacement creates a polarization gradient that forms a confining potential, localizing and laterally separating electrons and holes within the moir\'e supercell.  Excitons formed from flat band electrons and holes should therefore have not only the out-of-plane dipole moment familiar from MoSe$_2$/WSe$_2$ interlayer excitons, but an in-plane dipole moment as well.
\end{abstract}

\maketitle

\section*{Main}

When two lattice-matched transition metal dichalcogenide (TMD) monolayers are stacked with an interlayer twist angle, a moir\'e pattern with a periodicity larger than the monolayer period emerges \cite{GorbachevSTEMSulfHet,LinSTMWSe2BL,LeroyNatPhys2020,BediakoSTEMhBNConstrain,LeroyNatPhys2020,LeRoyTMDSTM,LuicanMayerSTMDFTWS2BL,SuenagaCrossoverAngleWSe2BL,PasupathySTM,AndreiSTM2023,Rosenberger2020moire}.  At very small angles ($\lesssim 1.5^\circ$) near the 0$^\circ$- or 60$^\circ$-aligned stacking, there is significant atomic reconstruction, leading to a network of commensurately stacked domains separated by domain walls and nodes \cite{GorbachevSTEMSulfHet,PasupathySTM,AndreiSTM2023,LeRoyTMDSTM,CarrPRBContModel,FalkoPiezoPRL,Heine2023}.  At twist angles larger than about 10$^\circ$, there is still a moir\'e super-period, but there is very little atomic reconstruction.  In between these two regimes, at angles around $2^\circ-5^\circ$, atomic displacements lead to moderate reconstruction, but sharp domain walls do not appear \cite{LeroyNatPhys2020,LinSTMWSe2BL,LuicanMayerSTMDFTWS2BL,SuenagaCrossoverAngleWSe2BL}.  Along with local interlayer stacking order, atomic displacements contribute to electronic state localization and the formation of flat bands \cite{NaikPRL2018,ZhanPRBMoS2BL_TB,NaikMoS2BL60,LinSTMWSe2BL,NaikPRBWSe2BL,LuicanMayerSTMDFTWS2BL}, which are in turn connected to the emergence of correlated phases like Mott insulators \cite{MakMott2020,CrommieWangMott} and superconductors \cite{MakWSe2BLsc2025,DeanWSe2BLFlatBands}. Previous theoretical studies have shown that the degree to which atomic displacement drives the localization of states depends on the details of the bands of the aligned bilayer structures \cite{NaikMoS2BL60,NaikPRBWSe2BL}. However, most of these studies have focused on twisted homobilayers. 

The twisted heterobilayer MoSe$_2$/WSe$_2$ has attracted attention as a platform for interesting exciton physics \cite{XXuMWseExciton2015,WurstbauerILE2017,KornILE2017}. Because of its Type II band alignment, it hosts long-lived interlayer excitons, which may enable photovoltaic and light-emitting diode technologies \cite{XXuMWseExciton2015}. However, MoSe$_2$/WSe$_2$ bilayers often have an interlayer twist \cite{HogeleMoWSeMoireExciton,XXuMoireExciton2019,XXuMoireTrion2021,MatsudaExciton2023,DeotareMoireExciton2021,WieczorekExcitonNatComm2023}, so it is important to understand how the resulting moiré physics, including the emergence of flat bands and correlated phases, influences exciton behavior.   Existing theoretical studies of MoSe$_2$/WSe$_2$ have provided insight via continuum relaxation models and electronic calculations of strained aligned bilayers \cite{FalkoQDs2022,FalkoTwirl2023,FalkoTwirl2,FalkoPEFE2021}, molecular dynamics (MD) relaxations and tight-binding models \cite{LischnerTB}, and machine-learning assisted MD and density functional theory (DFT) models \cite{ChenMoWSeMLDFT}. 

In this work, we provide a completely \emph{ab initio} study of near-60$^\circ$ MoSe$_2$/WSe$_2$ bilayers, using density functional theory to calculate both the atomic relaxations and electronic structure.  We show that flat bands emerge in both the valence and conduction bands in the intermediate reconstruction regime, at around a 3$^\circ$ twist away from 60$^\circ$ alignment.  There is both interlayer and lateral separation of the electrons and holes associated with the flat bands, suggesting that excitons formed from these carriers are not only layer-indirect (with an out-of-plane dipole moment) but also have an in-plane dipole moment. We show that in-plane atomic displacements lead to a confining potential that laterally localizes electrons and holes, indicating that piezoelectric effects dominate the moir\'e physics in near-60$^\circ$ MoSe$_2$/WSe$_2$, even when reconstruction is incomplete.

\section*{Structural and Electronic Calculations}

 \begin{figure}
  \includegraphics[angle=0,width=0.99\columnwidth]{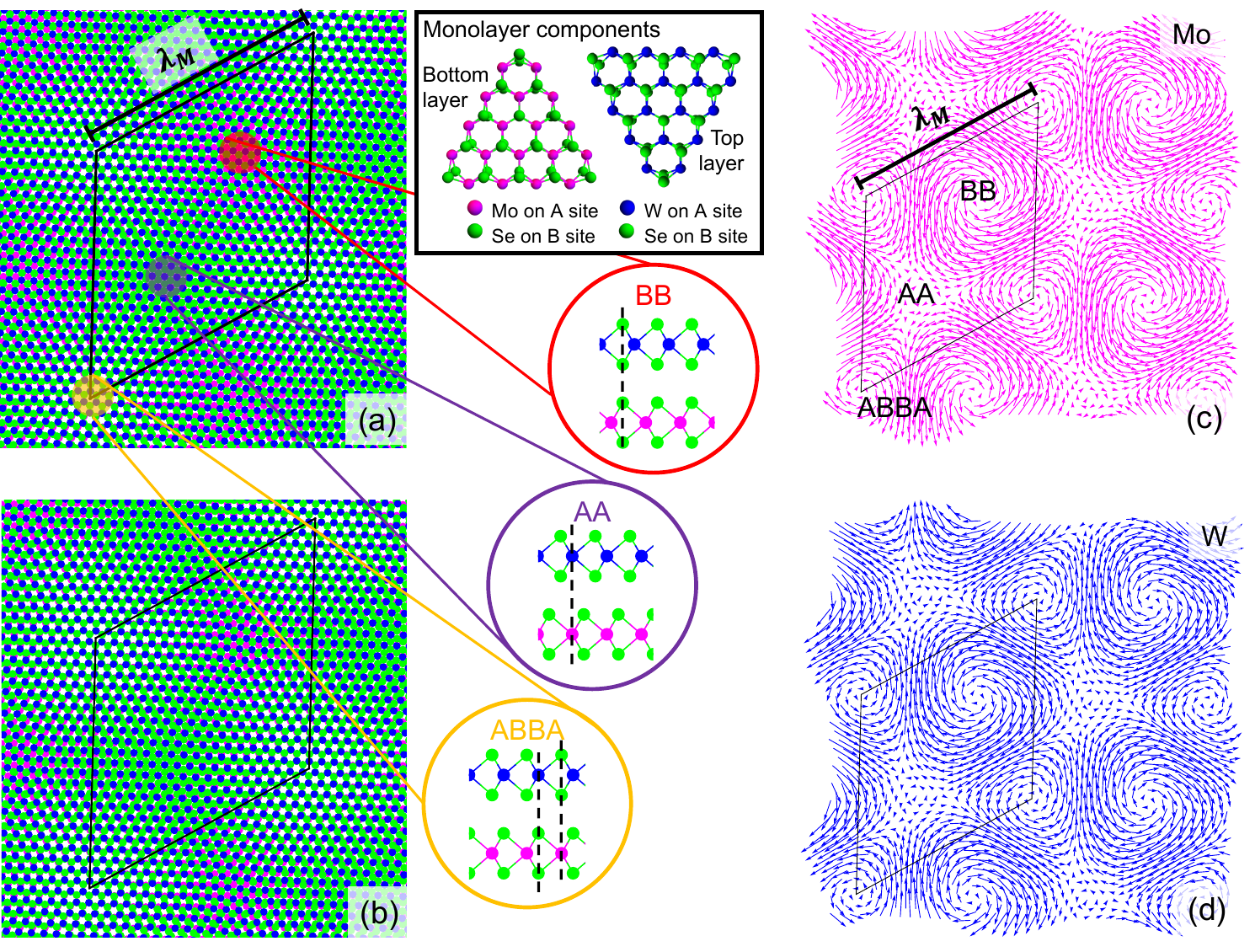}
  \caption{\label{fig:397Struct} Structural relaxation of 62.88$^\circ$ twisted MoSe$_2$/WSe$_2$ bilayer. (a) Unrelaxed and (b) relaxed structure of 62.88$^\circ$ twisted MoSe$_2$/WSe$_2$. Top inset shows the monolayer TMD components of the bilayer, and circular insets identify the regions of high symmetry stacking in the moir\'e supercell.  The moir\'e supercell length, $\lambda_M$, is 6.56 nm.  Panels (c, d) show the in-plane displacement fields of (c) the Mo atoms and (d) the W atoms. The displacement field is defined as $\mathbf{u}_k = \mathbf{r}^f_k - \mathbf{r}^0_k$, where $\mathbf{r}^0_k$ is the position of atom $k$ in the unrelaxed lattice, and $\mathbf{r}^f_k$ is the relaxed position of atom $k$. Selenium displacement fields are qualitatively the same and are shown in the SI \cite{SI}.) Atoms in the Mo and W layers relax with an opposing twist about the ABBA and BB high symmetry points, while the atoms near the AA high symmetry point relax relatively little.}  
\end{figure} 

Using density functional theory implemented in VASP \cite{VASP}, we compute the structural relaxation and the electronic bands of a variety of near-$60^\circ$ MoSe$_2$/WSe$_2$ heterobilayers with twist angles as small as 62.88$^\circ$. (See Methods section for full calculation details.)  Results of the structural relaxation for the 62.88$^\circ$ system are shown in Figure 1.  In the unrelaxed twisted structure (Fig. 1a), high symmetry stackings appear at three points in the moir\'e supercell.  Using the convention that the metals reside at the A sites of the hexagonal lattice and the chalcogens at the B sites, the high symmetry stackings are BB, where the chalcogens in the two TMD layers are aligned, AA, where the metals are aligned, and ABBA, where metals in each layer are aligned with chalcogens in the other layer.  
ABBA is frequently referred to as 2H stacking. 

The twisted bilayer relaxes to the structure shown in Fig.~1b, where the regions with ABBA stacking have grown, regions with near-BB stacking have shrunk, and AA-stacked regions have become more sharply triangular.  Reconstruction into domains of aligned bilayer has begun, but it is incomplete. The atoms achieve these new in-plane positions via opposing rotations about the ABBA and BB points (Fig 1c, d).  The atoms in the MoSe$_2$ layer twist in a counter-clockwise sense about the BB point and in a clockwise sense about the ABBA point, while the atoms in the WSe$_2$ layer exhibit the opposite twisting behavior.   Our relaxed structures agrees with STM experiments on MoSe$_2$/WSe$_2$ bilayers \cite{PasupathySTM} and theoretical calculations for generic TMDs showing that the lattice prefers to maximize the most energetically favorable stacking configuration, which is ABBA, and minimize the energetically less favorable BB stacking \cite{CarrPRBContModel,NaikPRL2018}. The maximum magnitude of the displacement field in the 62.88$^\circ$ bilayer is about 0.2 Angstroms, with slightly larger magnitudes in the MoSe$_2$ layer as compared to the WSe$_2$ layer. (Displacement fields with the same qualitative structure are observed for angles as large as about a 7$^\circ$ twist away from 60$^\circ$, where maximum displacement norms are about 1/6 of the 62.88$^\circ$ value \cite{SI}.)

 \begin{figure}
  \includegraphics[angle=0,width=0.99\columnwidth]{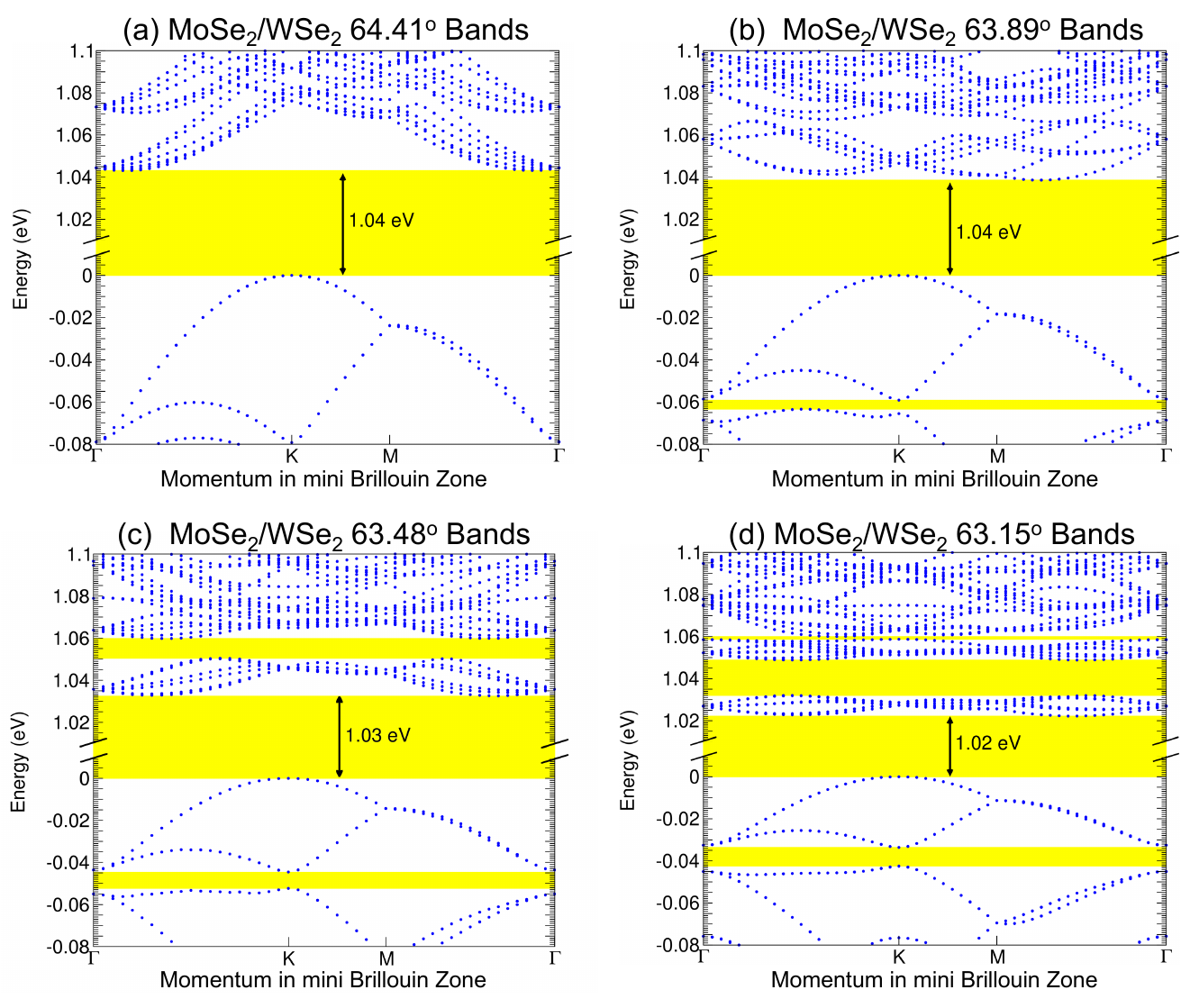}
  \caption{\label{fig:BandsWAngle} Emergence of flat bands with decreasing twist angle. Band structures of MoSe$_2$/WSe$_2$ bilayers at (a) 64.41$^\circ$, (b) 63.89$^\circ$, (c) 63.48$^\circ$, and (d) 63.15$^\circ$ plotted with respect to momentum in the mini-Brillouin zone. The valence band flat band emerges at 63.89$^\circ$, and the conduction band flat band emerges at 63.48$^\circ$.  As the twist angle decreases, the bandwidth of each flat band narrows.  Valence band flat bands are two-fold degenerate, while conduction band flat bands are six-fold degenerate. Bands are considered ``flat" if they are separated from the continuum of valence or conduction bands by a gap that spans the entire mini-Brillouin zone.}
\end{figure} 

After structural relaxation, we compute band structures for the near-60$^\circ$ twisted MoSe$_2$/WSe$_2$ bilayers.  We include spin-orbit coupling in our band structure calculations to capture the significant spin-orbit splitting in the TMDs, especially evident at the valence band edges at the K-point in the Brillouin zone of the aligned system.  Figure 2 shows the band structures of systems over the critical range of angles in which the flat bands emerge.  At a 64.41$^\circ$ twist, there are no flat bands in either valence or conduction bands.  At 63.89$^\circ$, a twofold degenerate flat band emerges in the valence band, with a bandwidth of about 59 meV.  At 63.48$^\circ$ twist, a sixfold degenerate conduction band flat band emerges, with a band width of about 18 meV, and the valence band flat band narrows.  At 63.15$^\circ$, both valence and conduction band flat bands continue to narrow and a second conduction band flat band emerges. The angles at which we see flat bands emerging are similar to the 3$^\circ$-5$^\circ$ critical angles observed and predicted in other twisted TMD systems \cite{LinSTMWSe2BL,NaikPRBWSe2BL,Heine2023,ChenMoWSeMLDFT}.  Taking our relaxation and band structure results together, we note that flat bands, along with the potential for physics dominated by electron correlations, emerge in the moderate reconstruction regime, well before the complete atomic reconstruction into domains of aligned stacking. 

To further understand the moir\'e physics in the MoSe$_2$/WSe$_2$ system, we look at the localization of the states associated with the flat bands. Figure 3 shows the modulus squared of the wavefunctions associated with the highest valence and lowest conduction band flat bands in the 62.88$^\circ$ system. Across the moiré Brillouin zone (mBZ)—sometimes called the mini Brillouin zone, since it is much smaller than the Brillouin zone of the aligned bilayer—the conduction band flat band states are localized strongly in the AA-stacked region of the moiré supercell, with weight in both the molybdenum and the tungsten layers.  In the valence band flat band, the state is localized in the ABBA region at the mBZ Gamma point, and as we move across the mBZ, the state spreads out to other regions of the moiré supercell but always avoids the AA region. Weight in the valence band flat band is only in the tungsten layer. Across Fig. 3, only weight on the metal atoms is shown, since weight on the selenium layers is an order of magnitude smaller \cite{SI}.  We emphasize that the electrons and the holes in the ~63$^\circ$ twisted MoSe$_2$/WSe$_2$ bilayer are localized in different regions in the moiré supercell, i.e. they have a physical lateral displacement in space.  Therefore, if excitons form from the carriers associated with the flat bands, they would have not only an out-of-plane dipole, since the holes are in the WSe$_2$ layer and the electrons have weight in both the MoSe$_2$ and WSe$_2$ layer, but also an in-plane dipole moment due to the lateral displacement. Such lateral separation could lead to the enhancement of exciton diffusion, as reported in WSe$_2$/WS$_2$ by Upadhyay et al \cite{HafeziX}.
 \begin{figure}
  \includegraphics[angle=0,width=0.9\columnwidth]{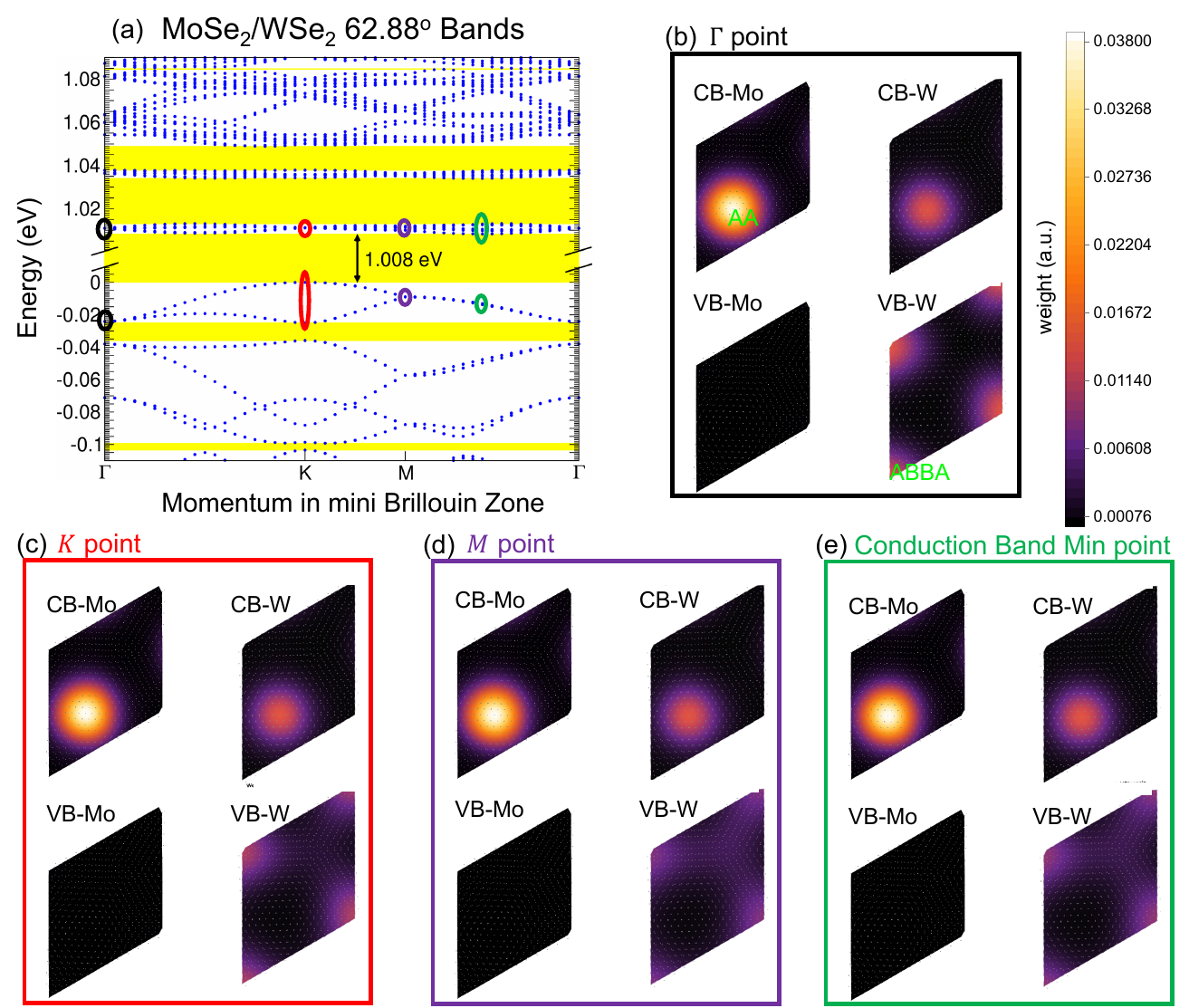}
  \caption{\label{fig:397Elec} 
Electronic bands and localization of flat band states for $62.88^\circ$ MoSe$_2$/WSe$_2$. (a) Electronic bands as a function of mini Brillouin zone momentum.  Both conduction band flat bands are six-fold degenerate.  The highest valence band flat band is two-fold degenerate.  Panels (b-e) show the modulus squared of the wavefunctions corresponding to the circled states in (a).  Each panel shows four moir\'e supercells: the top row shows the weight of the conduction band flat band state on the Mo (left) and W (right) atoms, and the bottom row shows the weight of the valence band flat band state on the Mo (left) and W (right) atoms. Conduction band weights are the sum of the six states that make up the conduction band flat band. Valence band weights are the sum of the two states that make up the valence band flat band. The conduction band states are localized around the AA-stacked region in the moir\'e supercell, while the valence band states are localized largely around the ABBA regions of the moir\'e supercell. Conduction band states have weight in both Mo and W layers, while valence band states have weight only in the W layer. The weight of the state on the Se atoms is an order of magnitude lower than the metal atom weights \cite{SI}.}
\end{figure} 

\section*{Effects of local interlayer stacking}

To understand the origin of the flat bands and carrier localizations, we first ask whether the twisted system inherits any properties from the regions of high-symmetry interlayer stacking. We find that the degeneracies and layer-localizations of both electron and hole flat bands can be traced back to the band edges of the aligned bilayers.  All of the aligned bilayer bands have the same qualitative features as the bands for the AA-stacked bilayer shown in Figure 4a: the valence band maximum is at K in the unit cell Brillouin zone and has weight only in the WSe$_2$ layer, while the conduction band minimum is at the Q point (i.e. between K and Gamma) with weight in both the MoSe$_2$ and WSe$_2$ layers. By simple band-folding arguments, we can map the valence band flat band to the K-point of the aligned band structures, since it has a two-fold degeneracy (from K and K’) and has weight only on the tungsten layer.  Similarly the conduction band flat band can be mapped to the aligned bilayer Q-points, since this flat band is sixfold degenerate, arising from the six Q-points in the aligned system (see inset of Fig. 4a) and has weight in both the molybdenum and tungsten layers.  Similar arguments have been made in studies of other TMD bilayers \cite{NaikMoS2BL60,Rubio_MoS2BL,NaikPRBWSe2BL}.  

These simple band-folding arguments do not, however, explain the lateral localization of the electrons and holes in the flat bands.  If we think of the twisted system as a patchwork of regions with different high symmetry stackings, then, looking at the band alignments of the three high symmetry stackings in Figure 4b, we would expect the holes to reside in the BB regions, though not by a wide margin of energy. The ABBA conduction band edge is about 50 meV lower than the conduction band edge in AA stacking, so we would expect the electrons to reside in the ABBA region, but our calculations show electrons strongly localized around the AA stacking region. It is clear that in the moderate reconstruction regime the emerging domains in MoSe$_2$/WSe$_2$ twisted bilayers cannot be straightforwardly modeled as regions of aligned stacking. 

 \begin{figure}
  \includegraphics[angle=0,width=0.99\columnwidth]{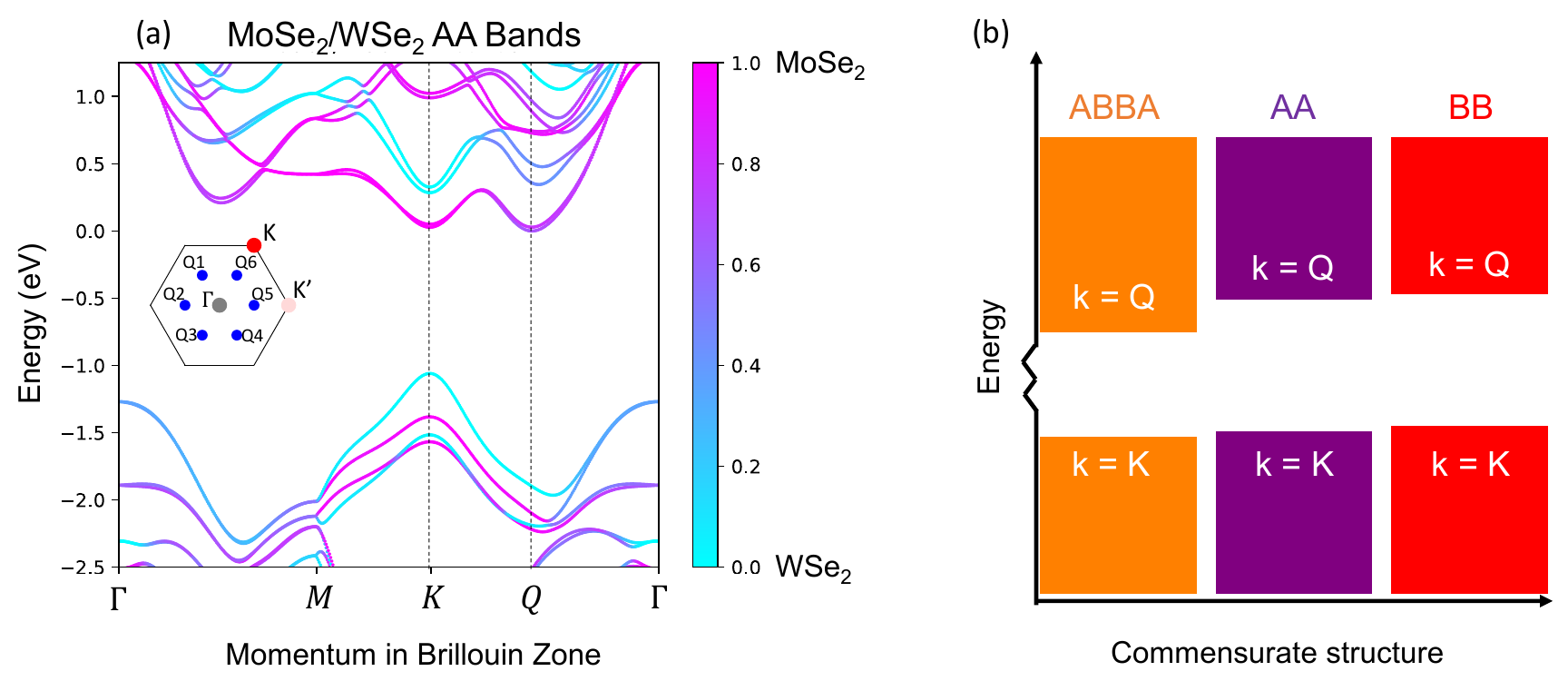}
  \caption{\label{fig:CommensurateBandData} Electronic structure of 60$^\circ$-aligned MoSe$_2$/WSe$_2$ bilayers. (a) Band structure of aligned AA-stacked MoSe$_2$/WSe$_2$. The conduction band minimum is at Q (between K and $\Gamma$ points), and the valence band maximum is at K. Inset: There are two K points in the system, K and K', and six Q points. In the twisted system, conduction and valence band flat band degeneracies and layer localizations are consistent with orgins in the band edges of the aligned system.  All high-symmetry stacking band structures (AA, ABBA, BB) have band edges at the same Brillouin zone points. (b) Band alignments of the valence band at K and the conduction band at Q for the three high-symmetry 60$^\circ$-aligned MoSe$_2$/WSe$_2$ bilayers.  If the twisted bilayer could be modeled as a patchwork of high-symmetry stacked regions, then electrons would be expected to reside in the ABBA region and holes in the BB region, but this is not what is observed in full twist calculations. The conduction band minimum (CBM) at Q in ABBA is 52 meV lower than the CBM at AA.  The valence band maximum (VBM) at K in BB is 9 meV higher than the VBM at AA and 18 meV higher than the VBM at ABBA.}
\end{figure} 

Previous work, mainly on TMD homobilayers, has shown that a combination of local stacking order and atomic displacement influences the state localization, with the degree of importance of each factor depending on specific material properties \cite{NaikMoS2BL60,NaikPRBWSe2BL}. For systems whose valence band (VB) flat bands originate from the K-valley in the aligned bilayers, hole localization is strongly influenced by atomic displacement rather than local stacking. The K-valley VB edges are formed by in-plane metal orbitals (almost entirely W d$_{xy}$ and d$_{x^2-y^2}$ with a small amount of Se p$_y$ and Se p$_x$ \cite{SI}), so the holes in these bands are largely insensitive to the specific stacking environment \cite{NaikPRBWSe2BL,ChenMoWSeMLDFT,FalkoAPL}, which agrees with our results.  The reason for the insensitivity of the conduction band electrons in our system to the interlayer stacking environment is less clear:  although the conduction band edges at Q in the aligned bilayers are dominated by in-plane orbitals, they also have a significant contribution from d$_{z^2}$ orbitals on the metal atoms \cite{SI}. Regardless, since our calculations indicate that the local stacking environment is not the primary driver of lateral state localization (Fig 4b), we consider the role of atomic displacements in both electron and hole localization \cite{FalkoPiezoPRL}.

\section*{Effects of atomic displacement}

To probe the role of atomic displacement, we consider the in-plane displacement field of the atoms, plotted for the metals in Figure 1c,d.  We neglect the z-component of atomic displacement because it results in an out-of-plane polarization that is, on average, several orders of magnitude smaller than the in-plane polarization \cite{SI}, and furthermore the out-of-plane displacements reflect the stacking environment \cite{SI}, whose contribution we have already considered. To compute the electronic effects arising from the atomic displacement, we fit the discrete displacement field we obtain from the DFT relaxation with a continuous function (see Methods for details).  Then we use this continuous displacement field function to compute the local polarization \cite{Vanderbilt2002}: $\bm{P}({\bf r}) = \frac{e}{\Omega_c} \sum_{\alpha=1}^3 Z^{\ast}_\alpha u_{\alpha}({\bf r})$, where $\Omega_c$ is the area of a monolayer unit cell, $\alpha$ runs over the atoms in a single monolayer unit cell (i.e. $\alpha =$ W, Se, Se in the WSe$_2$ layer), and $Z^{\ast}_\alpha$ is the Born effective charge of atom $\alpha$. (See Methods.)

The results of the polarization calculations are plotted in Figure 5a for the MoSe$_2$ and WSe$_2$ layers.  Both layers show a polarization field pointing towards the AA region, where the electrons associated with the conduction band flat band are localized.  To understand how this polarization field might also effect hole localization, we used it to compute the polarization-induced charge density, $\rho_{pol}=-\nabla \cdot \vec{P}$.  The results plotted in Figure 5b show a negative charge density at the ABBA regions of the bilayer and a positive charge density in the AA regions of the bilayer.  The areas of negative induced charge can be understood to attract holes, consistent with the ABBA-localization of holes in the valence band flat band, and the areas of positive charge density can be understood to attract electrons, consistent with the AA-localization of electrons in the conduction band flat band. Finally, we compute the potential landscape that arises from atomic displacements (Fig. 5c) by solving the Poisson equation with the induced charge density, and we find a potential maximum which traps electrons at AA and potential minima at ABBA and BB, where the holes are localized in our calculations. Potential wells are on the order of 100meV, which is consistent with calculations on WSe$_2$ homobilayers \cite{NaikPRBWSe2BL} and MoSe$_2$/WSe$_2$ experiments \cite{PasupathySTM}. Potential well depths increase with decreasing twist angle \cite{SI}. Notably, the electron potential well is deeper than that of the holes, even though the valence bland flat band emerges at slightly larger twist angles.  It's possible this is necessary to overcome both the large conduction band offset between the AA and ABBA region and the increased sensitivity of the band edge electrons to the interlayer stacking environment due to the contribution from $d_{z^2}$ orbitals. 

 \begin{figure}
  \includegraphics[angle=0,width=0.99\columnwidth]{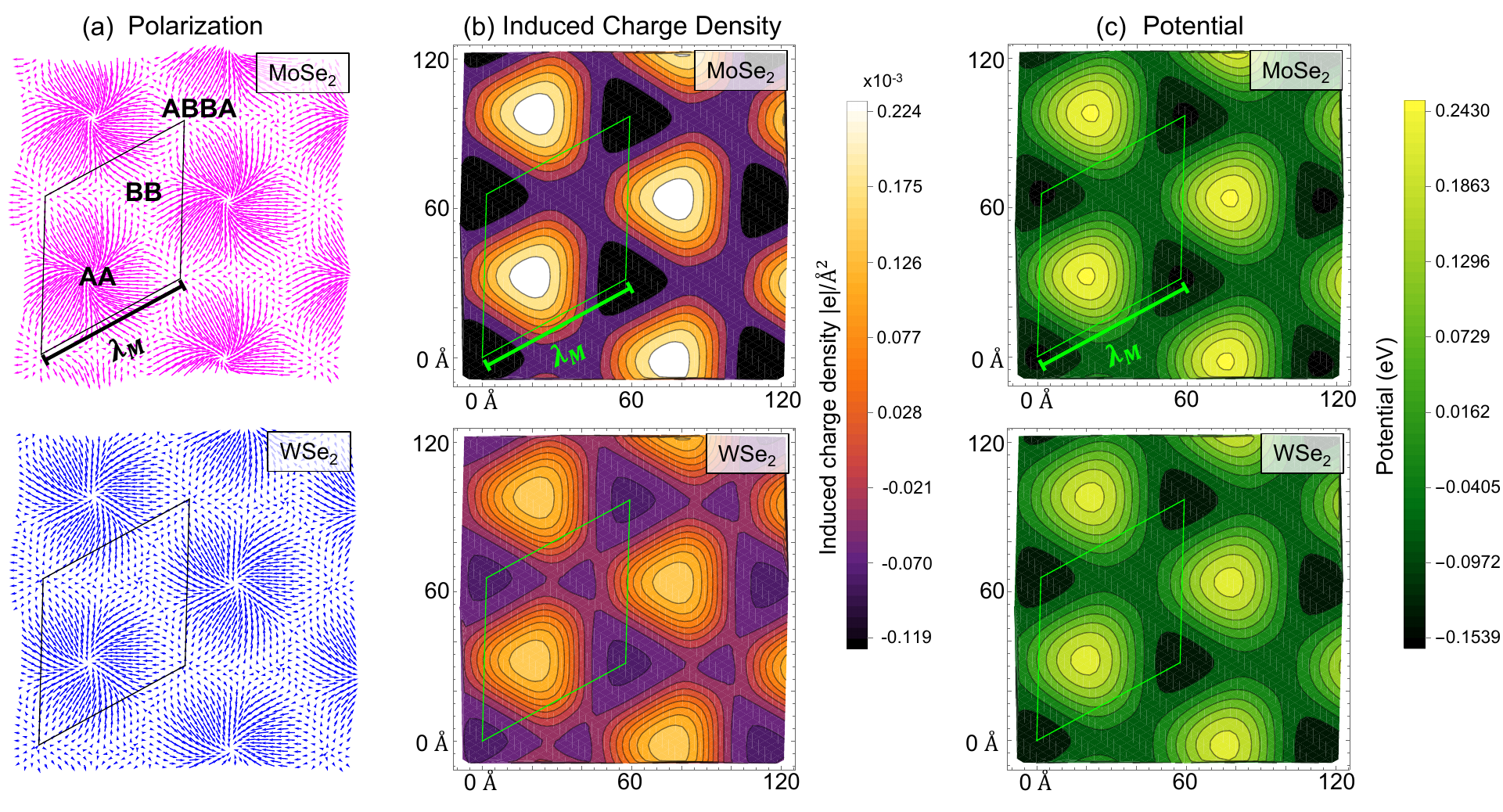}
  \caption{\label{fig:397uPolChg} In-plane atomic relaxation leads to induced charge density and moir\'e potential. (a) Polarization derived from the in-plane displacement field, $\vec{u}$, in 62.88$^\circ$ twisted MoSe$_2$/WSe$_2$ for the MoSe$_2$ layer (top) and the WSe$_2$ layer (bottom). $\lambda_M = 6.56$ nm in all panels. (b) Induced charge in (top) MoSe$_2$  and (bottom) WSe$_2$, computed from the negative divergence of the polarization. Induced charge is at a maximum at the AA-stacked regions and a minimum in the ABBA-stacked regions. (c) Potential landscape derived from the charge density induced by in-plane displacements, showing a maximum at the AA-stacked regions and a minimum at the ABBA-stacked regions, with a local minimum at the BB-stacked regions.  This is consistent with the localization of the electrons and holes plotted in Figure 3, where the electrons are drawn to the areas of positive potential and the holes drawn to the areas of negative potential.  Electron and hole lateral localization can thus be attributed mainly to piezoelectric effects in this heterobilayer. }
\end{figure}

\section*{Conclusions}

We present a completely \emph{ab initio} study of the structural and electronic properties of near-60$^\circ$ twisted MoSe$_2$/WSe$_2$ heterobilayers.   Flat bands emerge in both valence and conduction bands at about a 3$^\circ$ twist away from 60$^\circ$, and these bands get narrower as the twist angle approaches $60^\circ$.  The carriers associated with the flat bands inherit the layer-polarization from the aligned bilayer bands, with holes localized entirely in the WSe$_2$ layer and electrons hybridized between MoSe$_2$ and WSe$_2$ layers. However, piezoelectric effects drive lateral localization: polarization induced by the atomic displacements creates the potential that confines the flat band states within the moir\'e super cell. Electrons are confined at the AA-stacked regions, and holes are localized largely at the ABBA-stacked regions.  Thus excitons formed from carriers associated with the flat bands will have not only an out-of-plane dipole but an in-plane dipole moment as well.  

\section*{Methods}

\subsection*{First-principles calculations}
We use Density Functional Theory (DFT) implemented in VASP \cite{VASP}, with wavefunctions constructed using the Projector-augmented-wave (PAW) formalism \cite{BlochlPAW,KresseUltrasoftPAW}.  We approximate the exchange-correlation functional using the Perdew-Burke-Ernzerhof implementation of the Generalized Gradient Approximation \cite{PBE_GGA}.  The Mo, W, and Se potentials used each has six valence electrons.  We set the energy cutoff to 224.6 eV and relax the atoms until the residual forces are less than 0.6 meV/\AA\ .  The in-plane lattice constant is taken to be $a = 3.29$ \AA\ for both layers, and the out-of-plane lattice constant is $c = 22$ \AA, which yields about $12$\AA\ of vacuum between bilayer images. The van der Waals interaction is modeled using the DFT-D3 method of Grimme \cite{GrimmeDFTD3}.  Spin-orbit coupling is included in electronic calculations. The six-atom calculations of aligned MoSe$_2$/WSe$_2$ bilayers use an energy cutoff of 450 eV and an out-of-plane lattice constant of $c = 30$ \AA, but are otherwise the same as the moir\'e calculations.  A $9\times 9$ $\Gamma$-centered k-point mesh was used for the primitive cell calculations; the equivalent of a $6\times 6$ $\Gamma$-centered mesh was used for the supercells.  The moir\'e calculations for the smaller twist angles require large amounts of memory.  For the smaller angles, we computed each k-point in the band structure separately.

Twisted bilayers are periodic only for certain twist angles.
To construct the supercells used in this work, 
assume the primitive lattice vectors are ${\bf P}_\pm = (\frac{\sqrt{3}}{2},\pm\frac{1}{2})a$.
Two rotated supercells are created using different combinations of the primitive lattice vectors for integer $N$: 
\begin{equation}
\begin{array}{rclrccl}
{\bf S}_1 &=& (N+1){\bf P}_+ + N{\bf P}_- & ~&{\bf S}_2 &=&{\bf R}_{60^\circ}{\bf S}_1 \\
{\bf S}_1' &=& N{\bf P}_+ + (N+1){\bf P}_- & ~ &{\bf S}_2' &=&{\bf R}_{60^\circ}{\bf S}_1',
\end{array}
\end{equation}
where ${\bf R}$ is a rotation matrix.
The twisted bilayer is created by generating a supercell for the bottom layer using the ${\bf S}_j$ vectors
and a supercell for the top layer using the ${\bf S}_j'$ vectors.
The top layer is rotated to take the ${\bf S}_j'$ vectors into the ${\bf S}_j$ vectors \cite{NaikPRL2018}.

\subsection*{Moir\'e displacement, polarization, and potential calculation details}
TMD monolayers have no inversion symmetry and exhibit piezoelectricity.
To estimate the electrostatic potential induced by the atomic relaxation of the bilayers,
we use the local polarization\cite{Vanderbilt2002,WuVanderbilt06, BennettMeron23, GhosezDynChg98}  
\begin{equation}
{\rm \bf P}({\rm \bf r}) = \frac{e}{\Omega_c}\sum_\alpha {\rm \bf Z}^*_\alpha \cdot {\rm \bf u}_\alpha ({\rm \bf r}),
\label{polarization}
\end{equation}
where $ {\rm \bf u}_\alpha ({\rm \bf r})$ is the displacement field of the atoms in layer $\alpha$, the area of the unit cell is ${\Omega_c} = 9.37 {\rm \AA}^2$, and the ${\rm \bf Z}^*_\alpha$ are the Bohr effective charge tensors.
We computed the Bohr effective charge tensors in VASP using density functional perturbation theory \cite{Gonze97}.
For monolayers, we find

\begin{eqnarray}
{\rm \bf Z}_{\rm Mo}^* ({\rm MoSe_2}) &=& \left( \begin{array}{rrr}
-1.912  &  0.00  &   0.00\\
 0.00   & -1.912    & 0.00\\
 0.00    & 0.00    &-0.137
 \end{array}\right)\;\;\;
 {\rm\bf Z}_{\rm Se}^*({\rm MoSe_2})= \left( \begin{array}{rrr}
0.956  &  0.00  &   0.00\\
 0.00   & 0.956    & 0.00\\
 0.00    & 0.00    &0.068
 \end{array}\right)\\
 {\rm\bf Z}_{\rm W}^*({\rm WSe_2}) &=& \left( \begin{array}{rrr}
-1.318  &  0.00  &   0.00\\
 0.00   & -1.318    & 0.00\\
 0.00    & 0.00    &-0.114
 \end{array}\right)\;\;\;\;
 {\rm\bf Z}_{\rm Se}^* ({\rm WSe_2}) =  \left( \begin{array}{rrr}
0.659  &  0.00  &   0.00\\
 0.00   & 0.659   & 0.00\\
 0.00    & 0.00    &0.057
 \end{array}\right).
\end{eqnarray}
Notice the cations are effectively negative while the Se are positive \cite{PikeTMDZs17}.
The $xx$ and $yy$ components of ${\rm \bf Z}_{\rm Mo}^*$ for bilayers (both aligned and with large twist angles) are shown in Fig. S10 \cite{SI}.

We obtain the discrete displacement field from DFT, where the displacement of the $k_{th}$ atom is the relaxed position minus the unrelaxed position of atom $k$.  The unrelaxed in-plane positions of each atom are at the vertices of a hexagonal net with lattice vector $a=3.29$ \AA, and the unrelaxed out-of-plane separation between layers is the equilibrium separation of a 2H-stacked (ABBA) bilayer ($d_{W-Mo} = 6.53$ \AA\ ). We obtain displacements in (x, y, z) directions for each atom, but we only plot the in-plane displacements for each atom in Figure 1. To determine the displacement field as a continuous function, $\vec{u}(\vec{r})$, where $\vec{r}=(x, y, z)$, 
we fit the discrete displacements to a Fourier expansion\cite{FalkoPiezoPRL,MagorrianFalko21,MalicPRM24}
\begin{equation}
{\bf u}_\alpha ({\bf r}) =\sum_j {\bf u}^\alpha_j  \exp(i {\bf G}_j\cdot {\bf r})
\end{equation}
where the sum includes all reciprocal lattice vectors of the moir\'e cell, $\vec{G}$, with magnitude below a cutoff.
We vary the cutoff to minimize the reduced chi-square of the fit.
However, we do not exceed the 33rd reciprocal space star to avoid numerical instabilities.

The induced charge is given by
\begin{equation}
\rho_{\rm pol} = - \nabla \cdot {\rm \bf P},
\end{equation}
which we compute using the Fourier expansion of the displacement field, yielding
\begin{eqnarray}
\rho^{t/b} ({\bf r},z)
& = & \sum_j \rho^{t/b}_j \delta(z-z_{t/b})\exp(i {\bf G}_j\cdot {\bf r}),\label{charge}
\end{eqnarray}
where $\rho^t$ ($\rho^b$) is the charge induced by the polarization in the top (bottom) layer,
and we approximate the layers as being infinitely thin.
We determine the potential using Poisson's equation:
\begin{equation}
[\partial^2_{zz}+\nabla^2_{\bf r}]\varphi = - \left(\rho^t +\rho^b\right)/ \mathcal{E}_0,
\end{equation}
where $\mathcal{E}_0 = 55.263 \; e^2\, {\rm eV}^{-1} \mu {\rm m}^{-1}$.
Let's Fourier expand the potential generated by each layer as:
\begin{equation}
\varphi ^{t/b}({\bf r},z) =  \sum_j \Tilde \varphi^{t/b}_j(z) \exp(i {\bf G}_j\cdot {\bf r}).
\end{equation}
Then
\begin{equation}
[\partial^2_{zz}+\nabla^2_{\bf r}] \varphi ^{t/b}({\bf r},z) =  \sum_j \left(
\partial^2_{zz}\tilde \varphi^{t/b}_j(z) - {\rm G}_j^2\tilde \varphi^{t/b}_j(z)\right)  \exp(i {\bf G}_j\cdot {\bf r}),
\label{pot1}
\end{equation}
where ${\rm G}_j \coloneq |{\bf G}_j|$.
We expect the $z$ dependence of the potential will take the form:
\begin{equation}
\tilde \varphi^{t/b}_j(z) =  
\left\{ \begin{array}{ll}
\varphi^{t/b}_j \exp(-{\rm G}_j|z-z_{t/b}|)  & {\rm for} \; j\neq 0 \\
\varphi^{t/b}_0 |z-z_{t/b}| & {\rm for} \;j= 0
\end{array} \right.
\end{equation}
The $z$-derivatives are
\begin{eqnarray}
\partial^2_{zz} \exp(-{\rm G}_j|z-z_{t/b}|) & = & -2 {\rm G}_j \delta(z-z_{t/b}) +  {\rm G}_j^2 \exp(-{\rm G}_j|z-z_{t/b}|) \label{differentiate1}
\\
\partial^2_{zz} |z-z_{t/b}|  & = & 2 \delta(z-z_{t/b}),
\label{differentiate2}
\end{eqnarray}
and Eq.~(\ref{pot1}) becomes
\begin{equation}
[\partial^2_{zz}+\nabla^2_{\bf r}] \varphi ^{t/b}({\bf r},z) =  2 \delta(z-z_{t/b}) 
\left( \varphi^{t/b}_0- \sum_{j\neq 0} {\rm G}_j  \varphi^{t/b}_j \exp(i {\bf G}_j\cdot {\bf r}) \right),
\end{equation}
and the coefficients can be determined using Poisson's equation to be
\begin{equation}
\varphi^{t/b}_j = 
\left\{ \begin{array}{cl}
\rho^{t/b}_j /(2 {\rm G}_j  \mathcal{E}_0)& {\rm for} \; j\neq 0 \\
-\rho^{t/b}_0  /(2 \mathcal{E}_0)& {\rm for} \;j= 0
\end{array}. \right.
\end{equation}
The potential generated by each layer becomes
\begin{equation}
\varphi ^{t/b}({\bf r},z) =  
\frac{1}{2 \mathcal{E}_0}\left( -\rho^{t/b}_0 |z-z_{t/b}|+\sum_{j\neq 0} \frac{ \rho^{t/b}_j}{{\rm G}_j} \exp(-{\rm G}_j|z-z_{t/b}|) \exp(i {\bf G}_j\cdot {\bf r})\right).
\end{equation}
The total potential is simply the sum of the potentials from each layer:
\begin{eqnarray}
\varphi({\bf r},z) & = & -\frac{1}{2 \mathcal{E}_0}\left( \rho^{t}_0 |z-z_{t}| +  \rho^{b}_0 |z-z_{b}|  \right)
\\ & & +\frac{1}{2 \mathcal{E}_0} \sum_{j\neq 0}\frac{  \exp(i {\bf G}_j\cdot {\bf r})}{{\rm G}_j}
\left( 
\rho^{t}_j\exp(-{\rm G}_j|z-z_{t}|)+
\rho^{b}_j\exp(-{\rm G}_j|z-z_{b}|)
\right).
\end{eqnarray}
For charge neutrality $\rho^{t}_0 =- \rho^{b}_0$, 
which ensures the potential generated by the $j=0$ terms are constant outside the layers.
When each layer is neutral, $\rho^{t}_0 = \rho^{b}_0 = 0$.
The potential in each layer is
\begin{eqnarray}
\varphi({\bf r},z=z_t) & = &  \frac{1}{2 \mathcal{E}_0} \left( -\rho^{b}_0 d({\bf r}) +\sum_j \frac{  \exp(i {\bf G}_j\cdot {\bf r})}{{\rm G}_j}
\left( 
\rho^{t}_j+
\rho^{b}_j\exp(-{\rm G}_j d({\bf r}))
\right)\right) \label{p_top}
\\
\varphi({\bf r},z=z_b) & = &  \frac{1}{2 \mathcal{E}_0} \left(- \rho^{t}_0 d({\bf r}) +\sum_j \frac{  \exp(i {\bf G}_j\cdot {\bf r})}{{\rm G}_j}
\left( 
\rho^{t}_j\exp(-{\rm G}_j d({\bf r}))+
\rho^{b}_j
\right)\right). \label{p_bot}
\end{eqnarray}
where the interlayer distance is $d({\bf r}) = z_t({\bf r})-z_b({\bf r})$ of the relaxed moir\'e system.

\section*{Acknowledgements}
The authors acknowledge helpful discussions with Stanislav Tsoi, Philip Kim, Darshana Wickramaratne, and Noam Bernstein.  M.P. and C.S.H. disclose support for the research of this work from the Office of the Under Secretary of Defense for Research and Engineering (OUSD R\&E) through the Laboratory University Collaboration Initiative (LUCI) program (FY23-25 award). C.S.H. discloses support for the research of this work from the Applied Research for the Advancement of S\&T Priorities (ARAP) program (FY24-26 award).  M.P. discloses support for the research of this work from Basic Research 6.1 funding at the Naval Research Laboratory through the Nanoscience Initiative (NSI) program (FY25-28 award).

\bibliography{moirebibnew}

\end{document}


\title{Supplement to Atomic displacements drive flat band formation and lateral electron and hole separation in near-60$^\circ$ twisted MoSe$_2$/WSe$_2$bilayers}
\author{Madeleine Phillips* and C. Stephen Hellberg}
        \affiliation{Naval Research Laboratory, Center for Computational Material Science, Washington DC 20375, USA }
        \affiliation{*Corresponding author email: madeleine.phillips2.civ@us.navy.mil}

\maketitle

\begin{figure}
  \includegraphics[angle=0,width=0.95\columnwidth]{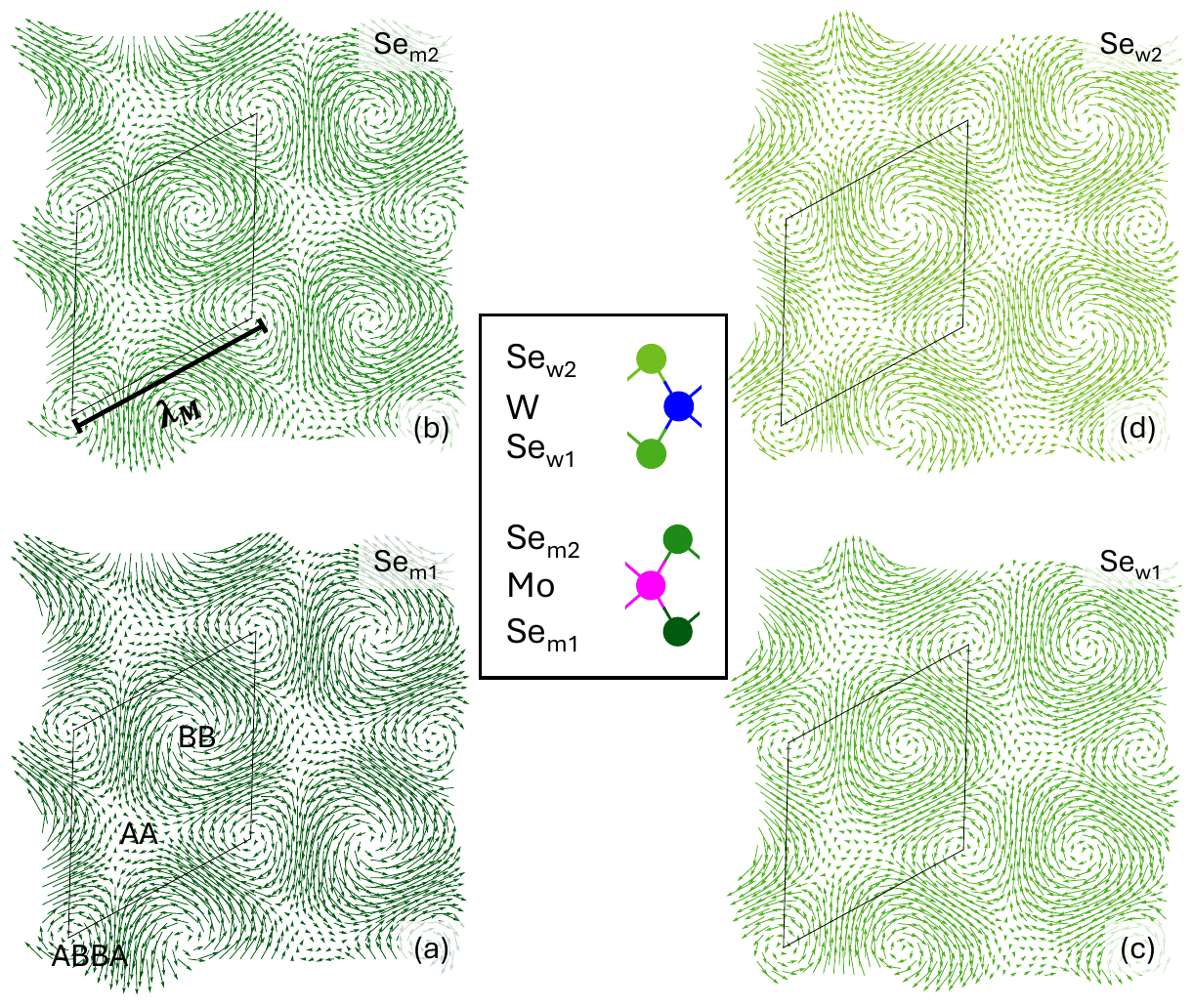}
  \caption{\label{SFIG_SeU} (a-d) Displacement fields for the selenium atoms in the $62.88^\circ$ twisted MoSe$_2$/WSe$_2$ bilayer.  The magnitudes of the selenum displacement fields are of the same order as the metal atom displacement fields shown in the main text.  The maximum norm of the displacement field in each layer, for Se$_{m1}$, Mo, Se$_{m2}$, Se$_{w1}$, W, Se$_{w2}$, is 0.209, 0.209, 0.213, 0.185, 0.182, 0.182 \AA, respectively. The inset shows a side view of the ABBA-aligned bilayer and labels the atoms in each layer. The moire length, $\lambda_M$, in all four panels is 6.56 nm. }
\end{figure}

\begin{figure}
  \includegraphics[angle=0,width=0.95\columnwidth]{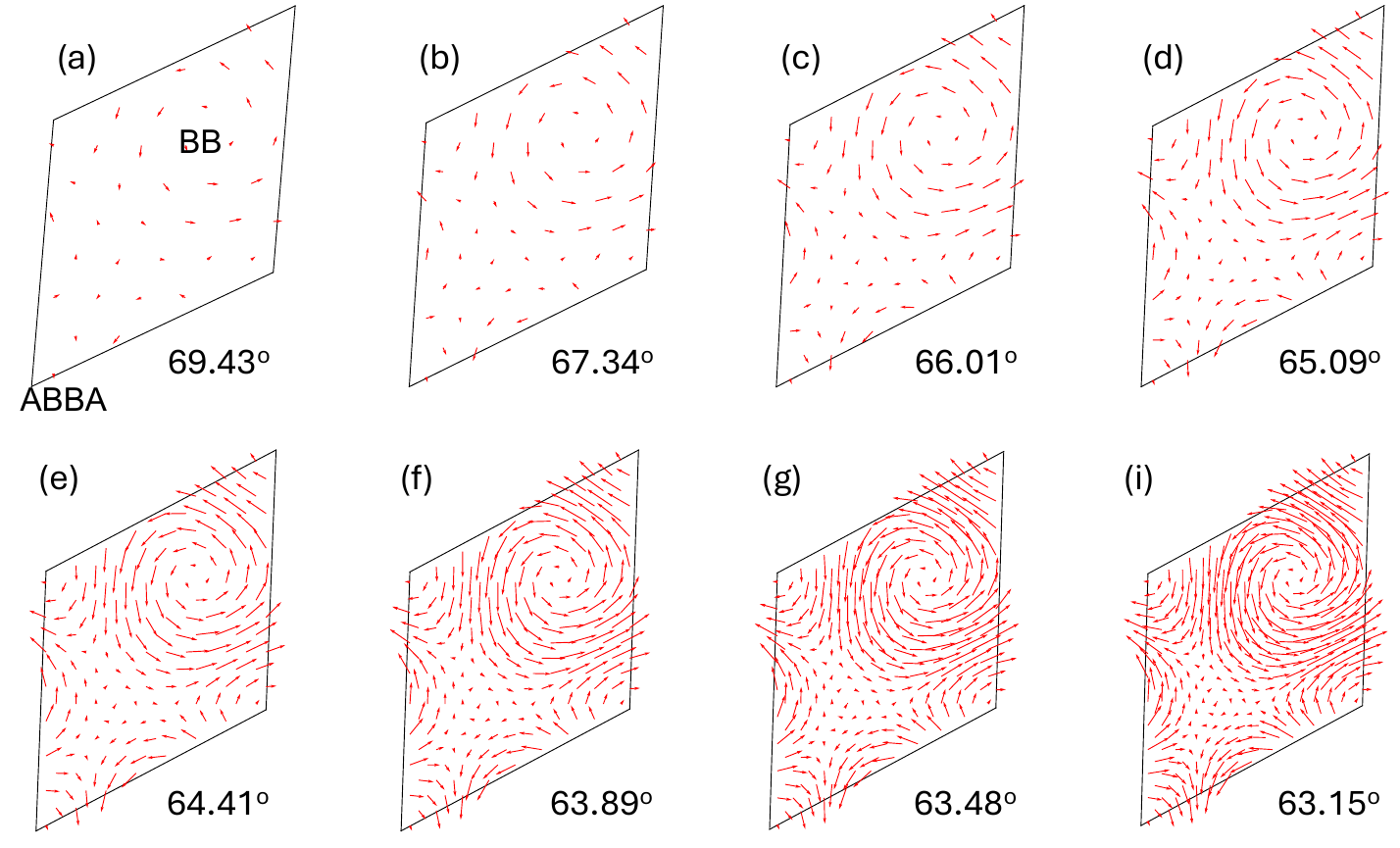}
  \caption{\label{SFIG_MoUwTH} In-plane displacement fields for the molybdenum atoms in twisted MoSe$_2$/WSe$_2$ bilayers ranging from $69.43^\circ$ to $63.15^\circ$. Atoms relax in a twisting motion about the BB point for all twist angles shown, but twisting about the ABBA point appears at angles smaller than about $7^\circ$ away from aligned $60^\circ$ stacking.}
\end{figure}

\begin{figure}
  \includegraphics[angle=0,width=0.95\columnwidth]{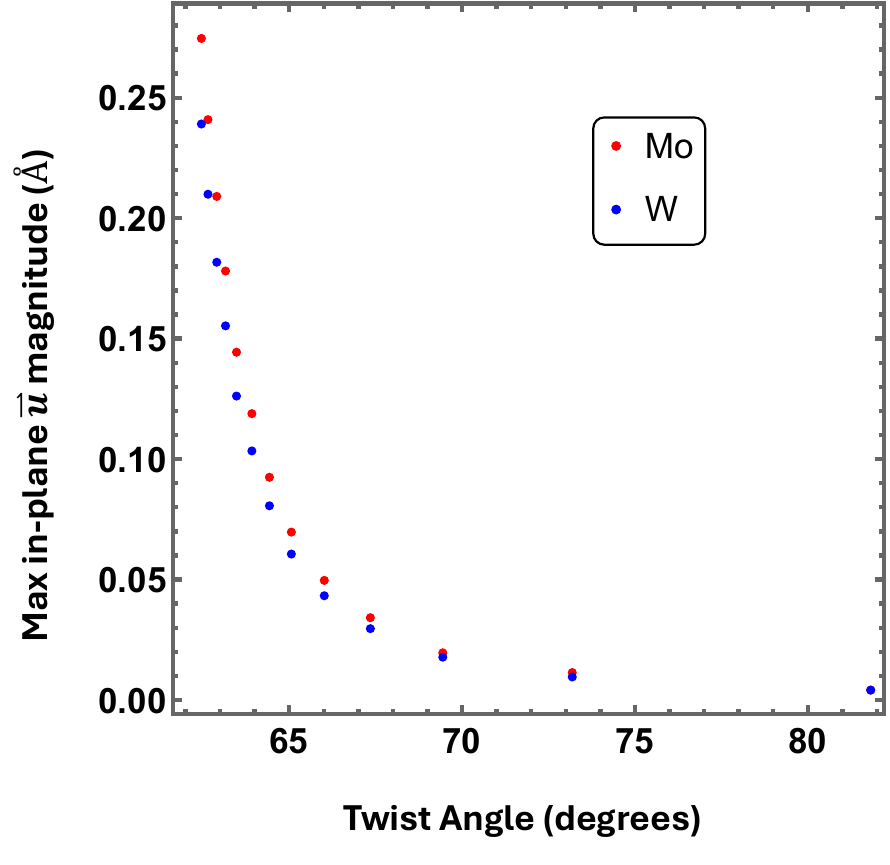}
  \caption{\label{SFIG_UMaxwTH} Maximum magnitude of the in-plane displacement field for molybdenum (red) and tungsten (blue) atoms in twisted MoSe$_2$/WSe$_2$ bilayers at a variety of twist angles.  As twist angle decreases towards $60^\circ$, the maximum displacement of atoms increases and the difference in maximum displacement for Mo vs. W layers increases.}
\end{figure}

\begin{figure}
  \includegraphics[angle=0,width=0.95\columnwidth]{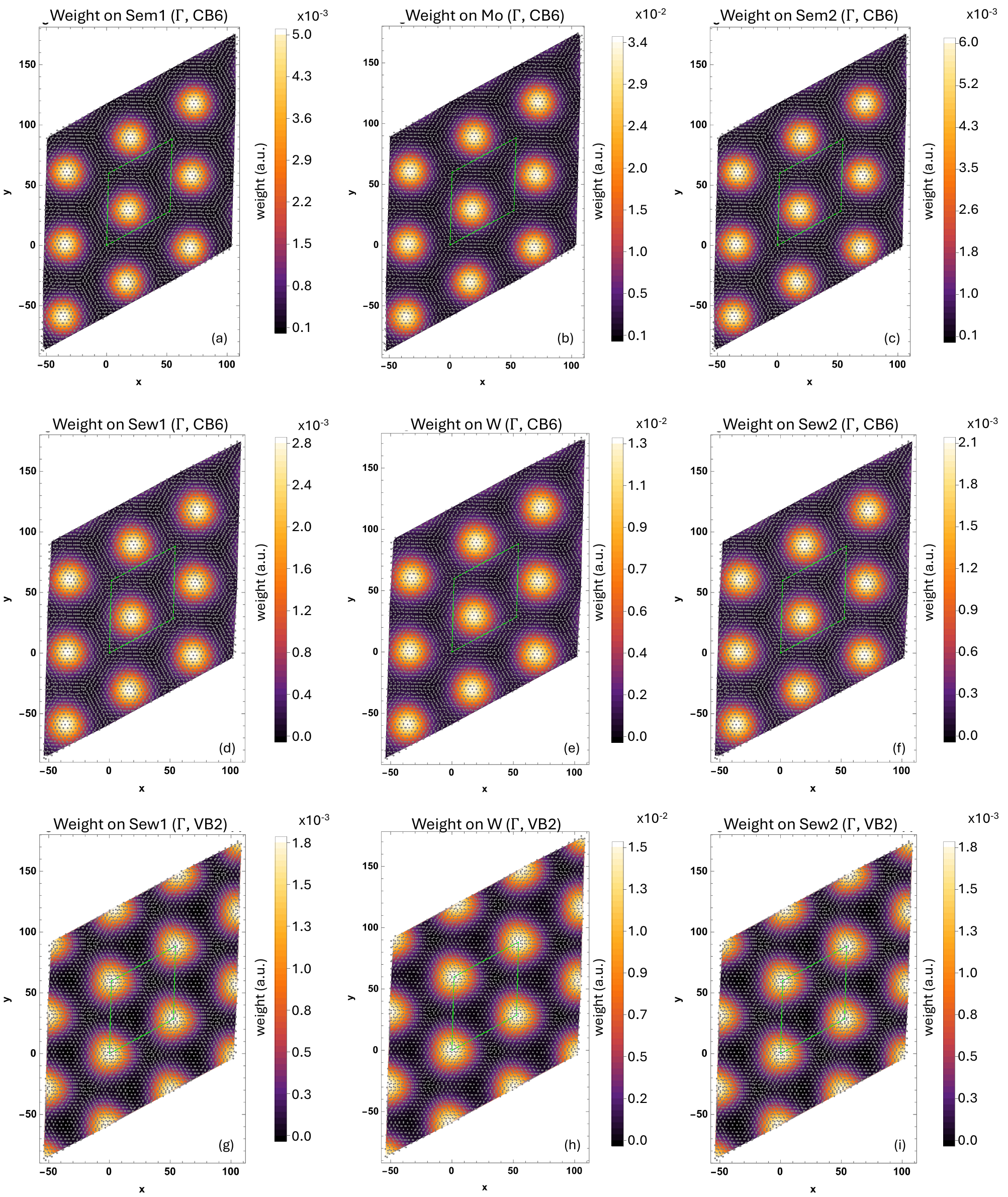}
  \caption{\label{SFIG_SeWeights331} Plots of the magnitude squared of the wavefunction associated with the conduction (a-f) and valence (g-i) band flat bands at the $\Gamma$ point in $63.15^\circ$ twisted MoSe$_2$/WSe$_2$. Atom labels Sem1, Mo, Sem2, Sew1, W, and Sew2 correspond to the inset of Figure S1. The conduction band flat band states plotted in (a-f) correspond to the sum of the weights of the lowest six conduction bands at the $\Gamma$ point.  The valence band flat band states plotted in (g-i) correspond to the sum of the weights of the highest two valence bands at the $\Gamma$ point. The weight of the states on the Se atoms is an order of magnitude less than the weight on the metal atom layers. Relative metal and selenium atom weights are expected to be qualitatively the same for the $62.88^\circ$ system. }
\end{figure}

\begin{figure}
  \includegraphics[angle=0,width=0.95\columnwidth]{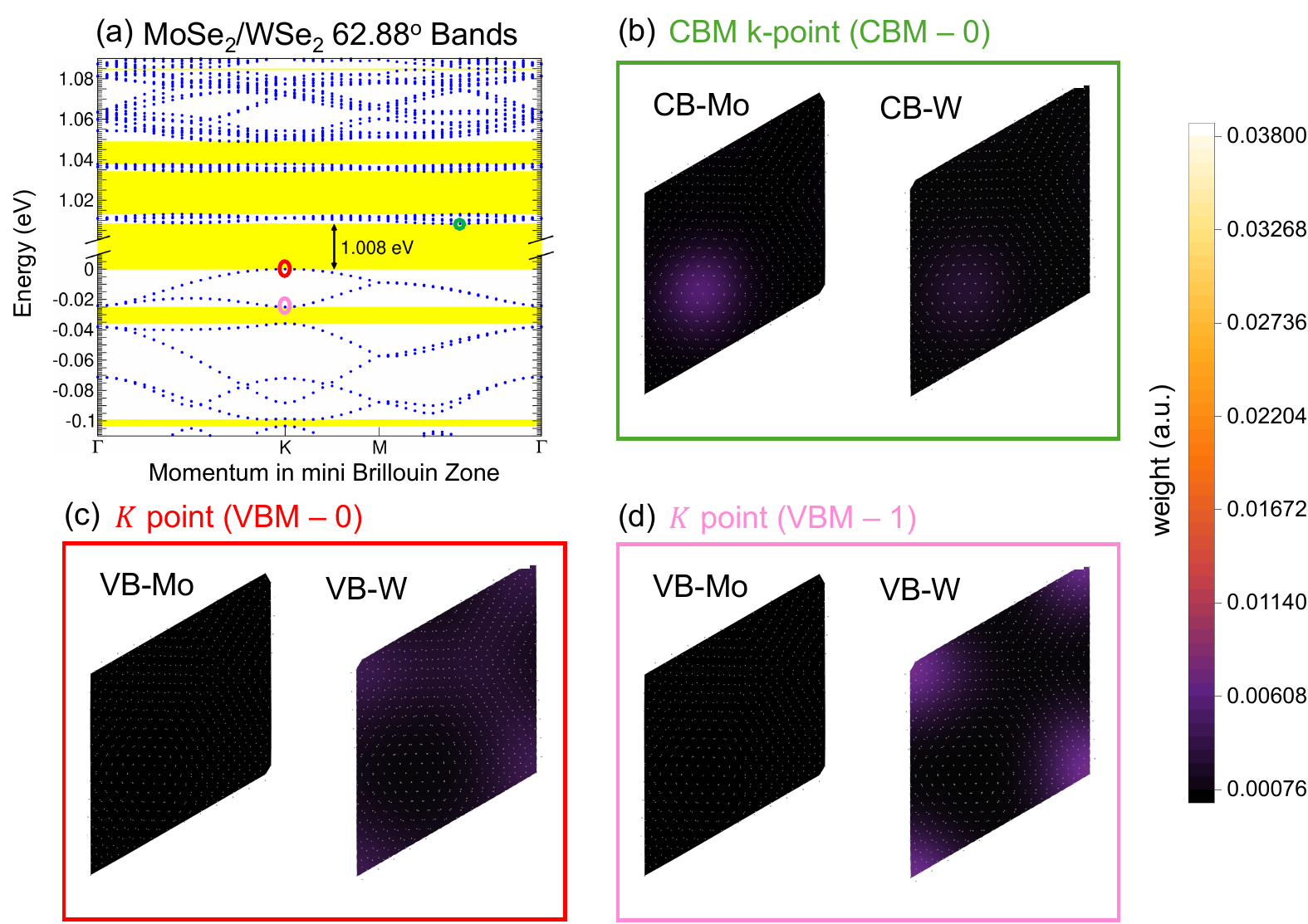}
  \caption{\label{SFIG_SoloBandEdges} (a) Band structure as a function of moir\'e Brillouin zone momentum for the $62.88^\circ$ twisted MoSe$_2$/WSe$_2$ bilayer.  (b-d) Modulus squared of the wavefunctions associated with the circled points in (a), with (b) showing the state at the conduction band (CB) minimum, (c) the state at the valence band (VB) maximum at K, and (d) the second-highest valence band state at K. The wavefunctions have the same localizations as those plotted in Figure 3 of the main text but with much smaller magnitudes.  The CB state has weight in both Mo and W layers at the AA-stacked region of the moire supercell.  The VB states (c, d) have weight only on the tungsten layer, with weight primarily at the ABBA-stacked corners of the moire supercell. The color scale is the same as that used in the main text. }
\end{figure}

\begin{figure}
  \includegraphics[angle=0,width=0.95\columnwidth]{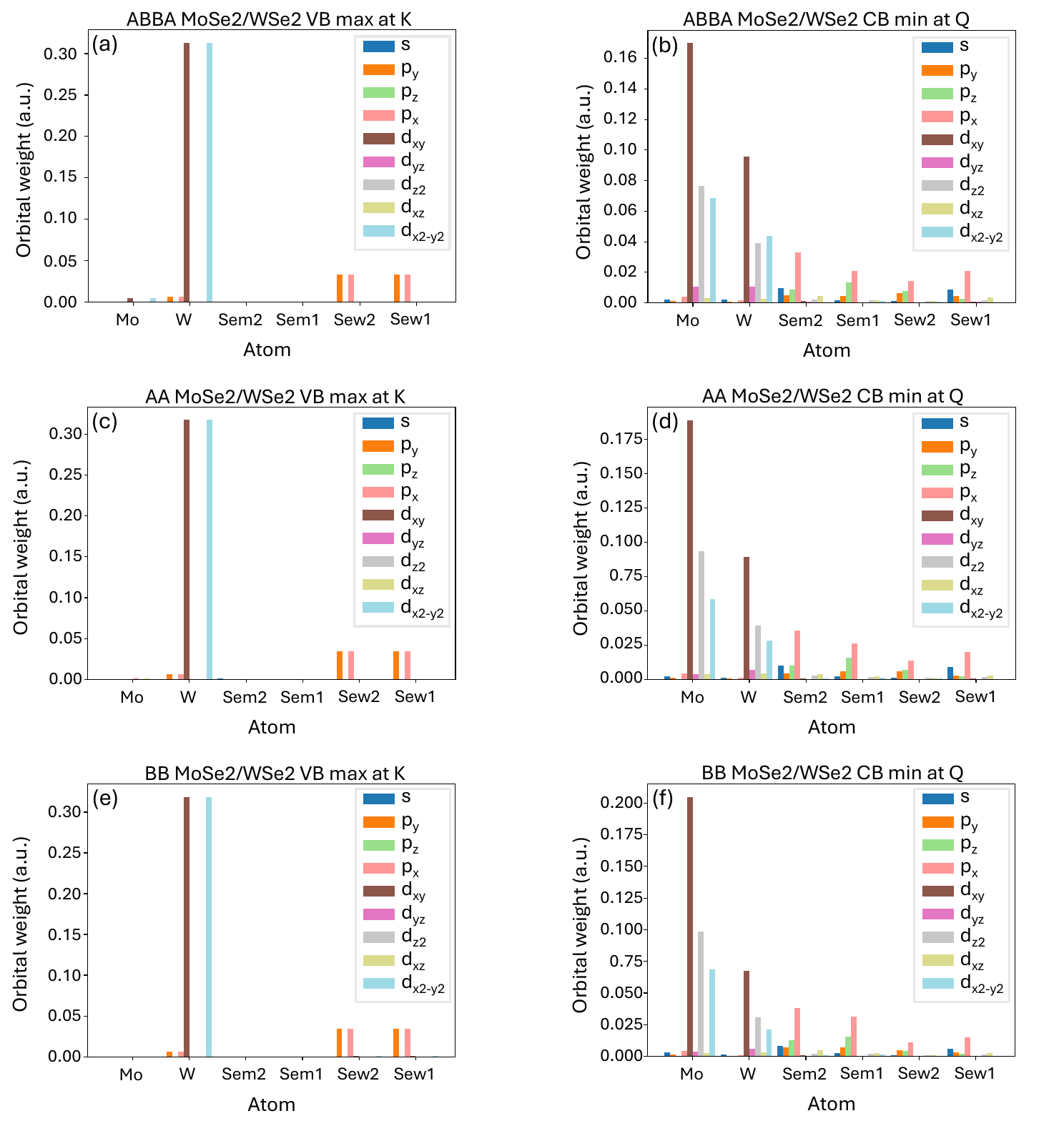}
  \caption{\label{SFIG_OrbWeights6atom} Orbital weights for states (a, c, e) at the valence band edge at K and (b, d, f) at the conduction band edge at Q for the $60^\circ$-aligned bilayers with (a-b) ABBA, (c-d) AA, and (e-f) BB stacking alignments.  The orbitals at the valence band edge are entirely in-plane (and mostly d) orbitals, while the orbitals at the conduction band edge are dominated by in-plane d-orbitals on the metal atoms but also show a significant contribution from d$_{z^2}$ orbitals on the metal atoms.}
\end{figure}

\begin{figure}
  \includegraphics[angle=0,width=0.95\columnwidth]{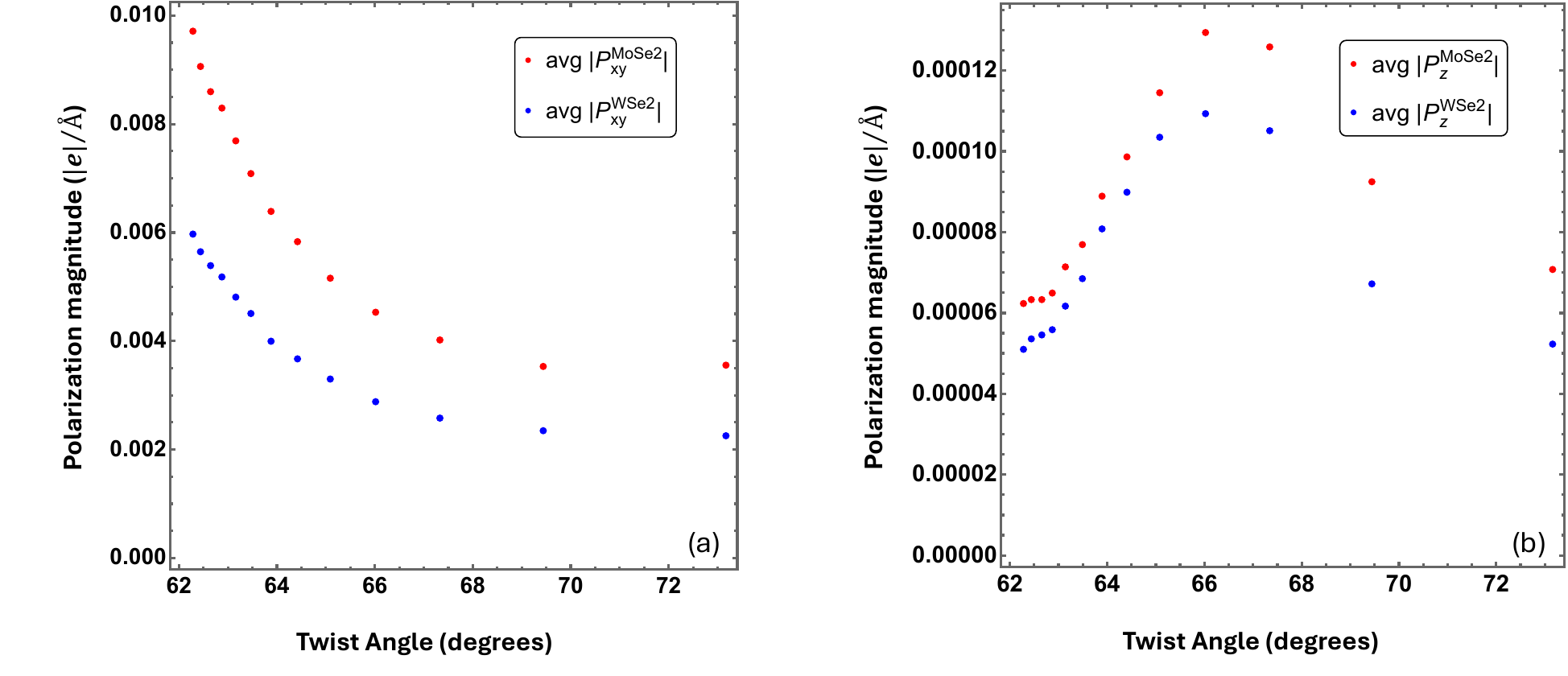}
  \caption{\label{SFIG_PAvgwTH} Magnitude of the average of (a) in-plane polarization and (b) out-of-plane polarization for MoSe$_2$ (red) and WSe$_2$ (blue) layers in twisted MoSe$_2$/WSe$_2$ bilayers at a variety of twist angles.  The magnitude of the in-plane polarization is one to two orders of magnitude larger than the magnitude of the out-of-plane polarization, so we neglect the out-of-plane component when we consider the piezoelectric effects in this heterobilayer. }
\end{figure}

\begin{figure}
  \includegraphics[angle=0,width=0.95\columnwidth]{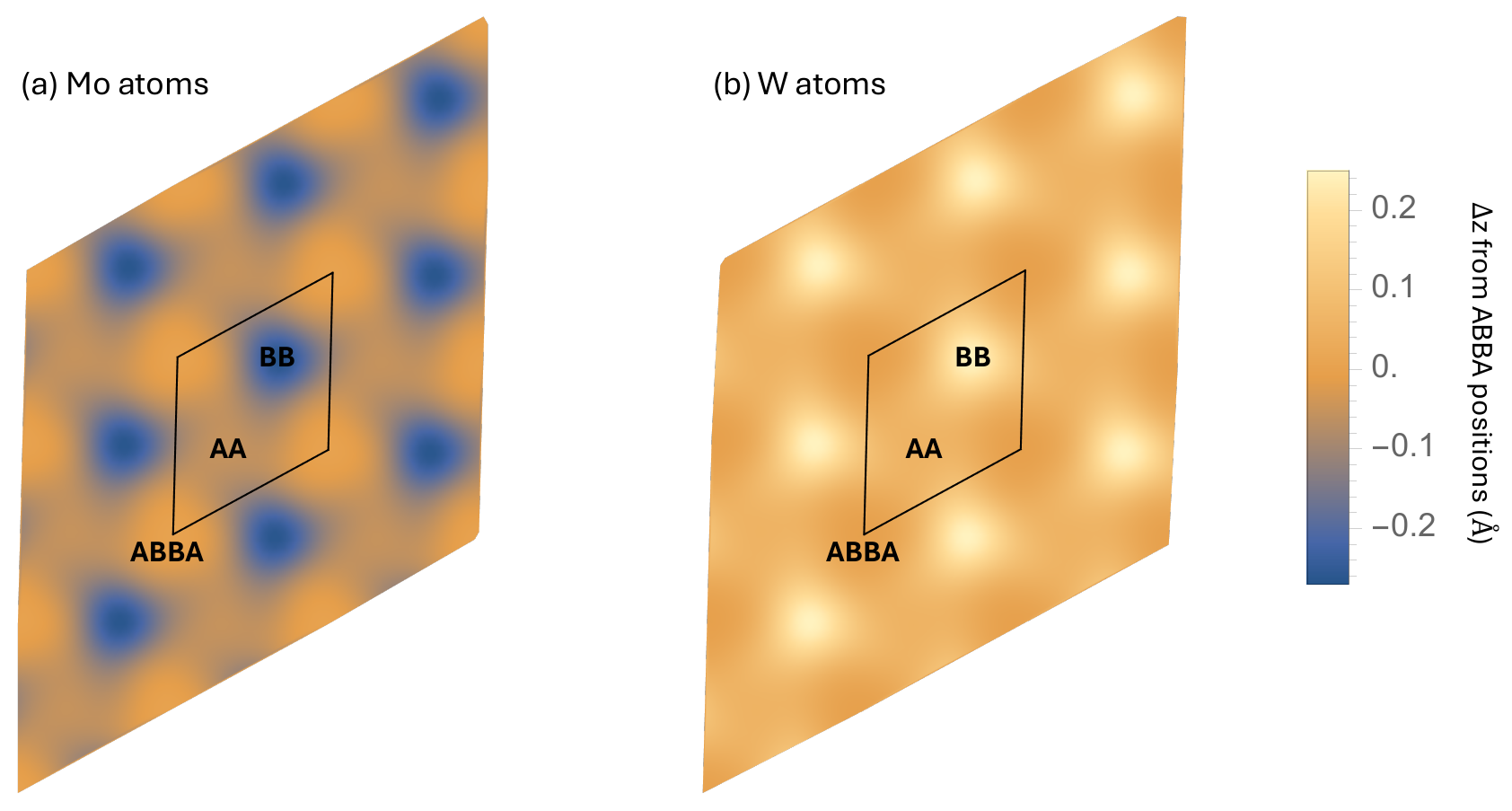}
  \caption{\label{SFIG_Zdiff397} Out-of-plane displacement of the (a) molybdenum and (b) tungsten atoms in the $62.88^\circ$-twisted MoSe$_2$/WSe$_2$ bilayer. Images show displacement from the ABBA equilibrium interlayer distance.  The greatest interlayer separation is around BB stacking, while the smallest interlayer separation is about the region of ABBA stacking. Recall that the MoSe$_2$ layer is below the WSe$_2$ layer.  The relaxed interlayer separation reflects the relative interlayer separation of the aligned ABBA, AA, and BB monolayers \cite{ILds_MoSe2WSe2}.}
\end{figure}

\begin{figure}
  \includegraphics[angle=0,width=0.95\columnwidth]{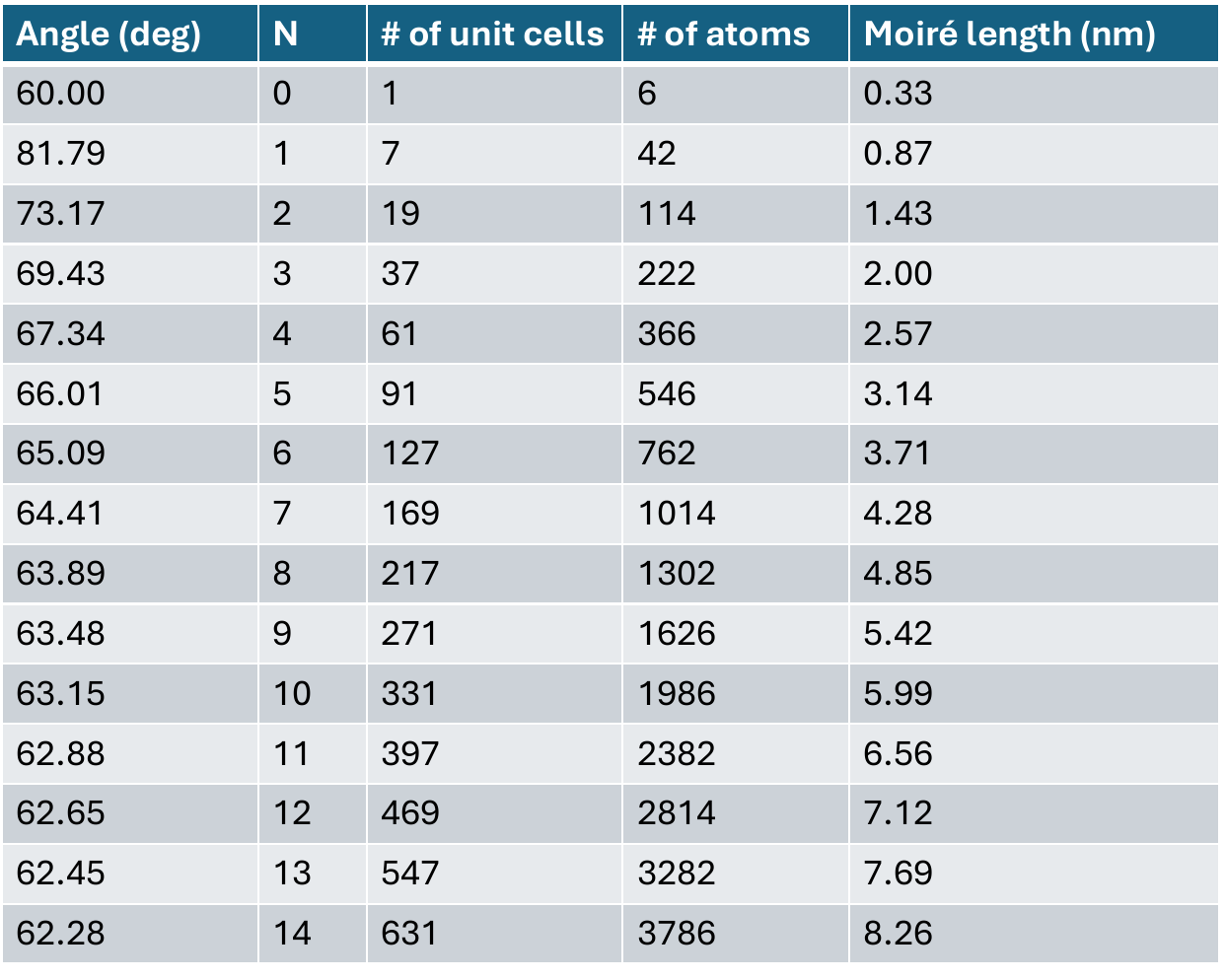}
  \caption{\label{SFIG_SCTable} Table showing characteristics of the commensurate twisted structures for which we have computed at least structural relaxation.  N indicates the variable in the expression $3N^2 + 3N + 1$, which gives the number of unit cells in a commensurately twisted structure.  The $62.88^\circ$ system, which is the smallest twist angle for which we computed both structural relaxation and electronic bands, contains 2382 atoms.}
\end{figure}

\begin{figure}
  \includegraphics[angle=0,width=0.95\columnwidth]{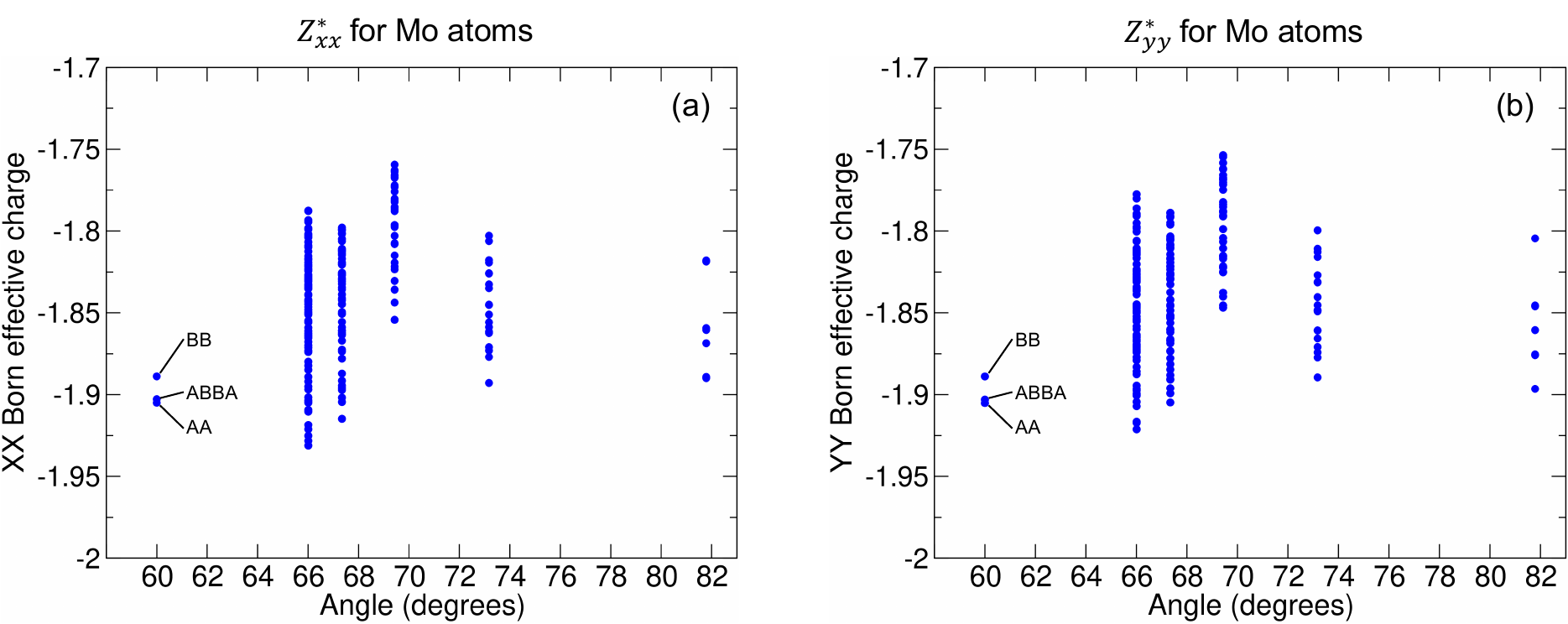}
  \caption{\label{SFIG_BornEffChgs} (a) The xx component and (b) the yy component of the Born effective charge tensor $Z^*$ for molybdenum atoms in MoSe$_2$/WSe$_2$ bilayers at several angles: $81.79^\circ$, $73.17^\circ$, $69.43^\circ$, $67.34^\circ$, and $66.01^\circ$, as well as in the aligned $60^\circ$ bilayers with ABBA, AA, and BB stacking. Due to computational constraints, we use the Born effective charge values for the aligned ABBA bilayer in our polarization calculations, since they lie within the range of the $Z^*_{xx}$ and $Z^*_{yy}$ values of the smallest twist angle systems for which we were able to compute $Z^*$.}
\end{figure}

\begin{figure}
  \includegraphics[angle=0,width=0.95\columnwidth]{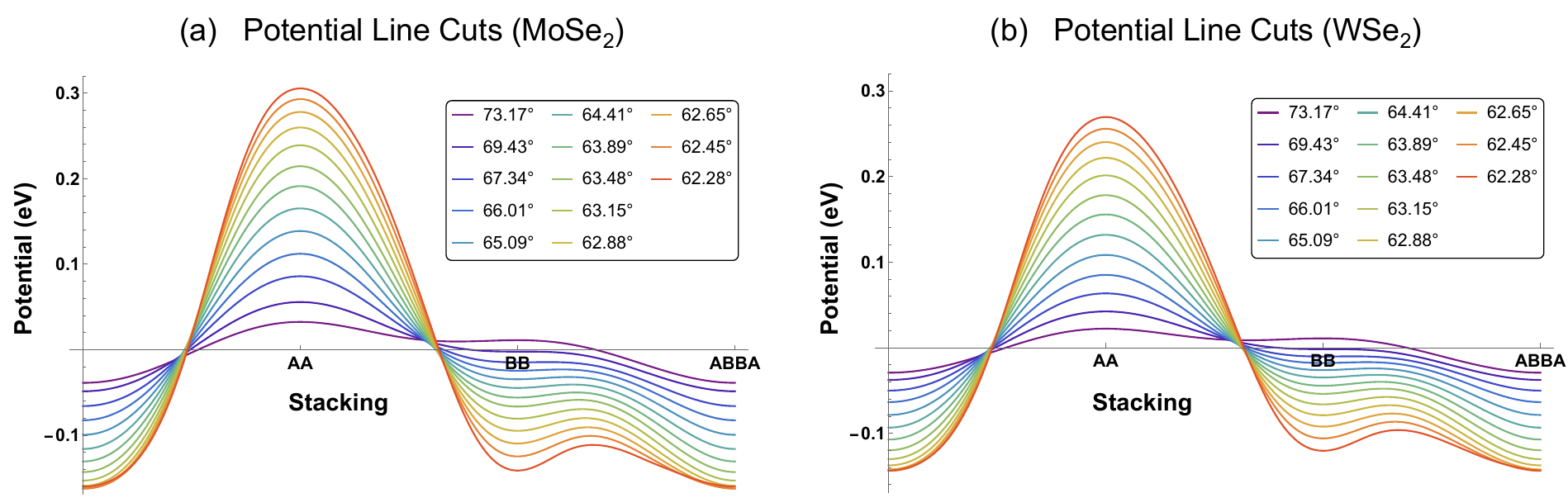}
  \caption{\label{SFIG_VLineCuts} Line cuts of the potential (as in main text Figure 5c) across the main diagonal of the moire cell for (a) the MoSe$_2$ layer and (b) the WSe$_2$ layer for systems with twist angles ranging from 73.17$^\circ$ to 62.28$^\circ$. Potential well depth increases as twist angle decreases. Note that we are able to relax structures with twist angles below 62.88$^\circ$ degrees, but computing bands for these smaller twist angles is prohibitively expensive in terms of computational cost.}
\end{figure}

\clearpage
\bibliography{moirebibnew}